\documentclass[graybox]{svmult}

\usepackage{mathptmx}       
\usepackage{helvet}         
\usepackage{courier}        
\usepackage{type1cm}        
\usepackage{makeidx}         
\usepackage{graphicx}        
\usepackage{multicol}        
\usepackage[bottom]{footmisc}

\makeindex             

\newcommand{\be}{\begin{equation}}
\newcommand{\ee}{\end{equation}}
\newcommand{\bea}{\begin{eqnarray}}
\newcommand{\eea}{\end{eqnarray}}
\newcommand{\lb}{\label}
\newcommand{\bdm}{\begin{displaymath}}
\newcommand{\edm}{\end{displaymath}}

\newcommand{\del}{\partial}

\newcommand{\ssst}{\scriptscriptstyle}

\newcommand{\intx}{\int\!\mathrm{d}^3x}

\newcommand{\X}{{\mathbf x}}

\begin{document}

\title*{Notes on semiclassical Weyl gravity}

\author{Claus Kiefer and Branislav Nikoli\'c}

\institute{Claus Kiefer \at Institute for Theoretical Physics,
  University of Cologne, Z\"ulpicher Strasse~77, 50937 K\"oln, Germany,
  \email{kiefer@thp.uni-koeln.de} 
\and Branislav Nikoli\'c \at 
Institute for Theoretical Physics,
  University of Cologne, Z\"ulpicher Strasse~77, 50937 K\"oln, Germany,
\email{nikolic@thp.uni-koeln.de}} 
%
%
\maketitle

\abstract{
In any quantum theory of gravity, it is of the
utmost importance to recover the limit of quantum theory
in an external spacetime. In quantum geometrodynamics 
(quantization of general relativity in the Schr\"odinger picture),
this leads in particular to the recovery of a semiclassical
(WKB) time which governs the dynamics of non-gravitational fields
in spacetime. Here, we first review this procedure with special
emphasis on conceptual issues. We then turn to an alternative
theory - Weyl (conformal) gravity, which is defined by a Lagrangian
that is proportional to the square of the Weyl tensor.
We present the canonical quantization of this theory and
develop its semiclassical approximation. We discuss in particular
the extent to which a semiclassical time can be recovered 
and contrast it with the situation in quantum geometrodynamics.
}

\section{Notes on semiclassical Einstein gravity}
\label{sec:1}

Among Paddy's many interests in physics was always the deep
desire to understand the relationship between classical and quantum
gravity. In his paper ``Notes on semiclassical gravity'', written
together with T.~P.~Singh in 1989, they write \cite{SP89}:
\begin{quote}
In the course of our investigation we came across a variety of methods
for defining classical and semiclassical limits, apparently different,
and all of which were possibly applicable to a quantum gravity. It
then became necessary to compare these methods and to settle, once and
for all, the relation of semiclassical gravity to quantum gravity.
\end{quote}
The understanding of semiclassical gravity was also a long-term
project by one of us, and we thus devote our festschrift contribution
to this topic. More precisely, the topic is the recovery of quantum
(field) theory in an external spacetime from 
canonical quantum gravity. We briefly
review the standard procedure of obtaining this limit from the
Wheeler-DeWitt equation of quantum general relativity (quantum
geometrodynamics). In the next two sections, we then apply these
methods to a different theory called Weyl gravity or conformal
gravity. This is the main concern of our paper. 

What is our motivation for doing so? Weyl gravity is a theory
without intrinsic scale. It seems therefore not appropriate, by itself,
to replace general relativity (GR) in the empirically tested
macroscopic limit. It may, however, be appropriate to serve as a
model for a fundamental conformally invariant theory, being of
relevance in quantum gravity and its application to the very early
universe. Many researchers entertain, in fact, the idea that Nature
does not contain any scale at the most fundamental level; see, for
example, \cite{Hooft15} and \cite{Barbour12}. In these following
sections, we shall outline the procedure for classical and quantum
canonical Weyl gravity and perform the semiclassical limit. We shall
point out in detail the similarities to and the differences from
quantum GR. We shall see, in particular, that while a semiclassical
time can be recovered, this time is of a different nature than the one
recovered from quantum GR. 

In canonical GR, the configuration variable is the three-metric
${h}_{ab}(\X)$, while the canonical momentum ${p}^{cd}(\X)$ is a
linear function of the extrinsic curvature (second fundamental form) 
$K_{cd}(\X)$. In the Dirac way of quantization, these variables are
heuristically transformed into operators acting on wave functionals,
\bea
\hat{h}_{ab}(\X)\Psi\left[h_{ab}(\X)\right] &=&
h_{ab}(\X)\cdot\Psi\left[h_{ab}(\X)\right]\ , \lb{5.4}\\
\hat{p}^{cd}(\X)\Psi\left[h_{ab}(\X)\right] &=& \frac{\hbar}{\I}
\frac{\delta}{\delta h_{cd}(\X)}\Psi\left[h_{ab}(\X)\right]\ . \lb{5.5}
\eea
The wave functionals are defined on the configuration space of all
three-metrics (plus non-gravitational fields, which are not indicated
here). In GR, one has four local constraints, the Hamiltonian
constraint and the three diffeomorphism (momentum) constraints. 
They are implemented in the quantum theory as restrictions on
physically allowed wave functionals \cite{OUP},
\bea
\hat{\mathcal H}_{\perp}^{\rm g}\Psi&:=&
\left(-16\pi G\hbar^2G_{abcd}\frac{\delta^2}{\delta h_{ab}\delta h_{cd}}
-\frac{\sqrt{h}}{16\pi G}(\,{}^{(3)}\!R-2\Lambda)\right)\Psi=0\ , \lb{5.21}\\
\hat{\mathcal H}_a^{\rm g}\Psi &:=& -2D_bh_{ac}\frac{\hbar}{\I}
\frac{\delta\Psi}{\delta h_{bc}} =0\ . \lb{5.22}
\eea
The quantum Hamiltonian constraint (\ref{5.21}) is called the Wheeler-DeWitt
equation. 
The momentum constraints (\ref{5.22}) guarantee that the wave
functional remains unchanged (apart possibly from a phase) under a
three-dimensional coordinate transformation. In the presence of
non-gravitational fields, we need the corresponding contributions 
$\hat{\mathcal H}_{\perp}^{\rm m}$ for (\ref{5.21}) and
$\hat{\mathcal H}_{a}^{\rm m}$ for (\ref{5.22}), see below.

The coefficents $G_{abcd}$ in front of the kinetic term in
(\ref{5.21}) are the components of the DeWitt metric, which is the
metric on configuration space. One of its important properties is its
indefinite nature. Using instead of $h_{ab}$ its scale part $\sqrt{h}$
(where $h$ denotes its determinant) and the conformal part
$\bar{h}_{ab}=h^{-1/3}h_{ab}$, the Wheeler-DeWitt equation reads
\bea
\lb{5.25}
\Big(6\pi G\hbar^2\sqrt{h}\frac{\delta^2}{\delta(\sqrt{h})^2} 
& - & \frac{16\pi G\hbar^2}{\sqrt{h}}\bar{h}_{ac}\bar{h}_{bd}
\frac{\delta^2}{\delta\bar{h}_{ab}\delta\bar{h}_{cd}}\nonumber\\
& - & \frac{\sqrt{h}}{16\pi G}(\,{}^{(3)}\!R-2\Lambda)\Big)
\Psi[\sqrt{h},\bar{h}_{ab}]=0\ .
\eea
One recognizes that the kinetic term connected with the local scale 
has a different sign. For this reason, the Wheeler-DeWitt equation is
of a (local) hyperbolic nature and $\sqrt{h}$ can be interpreted as a
local measure of {\em intrinsic time}. We shall introduce the scale
and conformal parts of the metric also for the Weyl theory below, but
as we shall see, the scale part (and thus the intrinsic time part)
will be absent in the Weyl version of the Wheeler-DeWitt equation.

An important step in understanding the semiclassical limit for the above
quantum equations is the WKB approximation \cite{SP89}. One starts
with the ansatz
\be
\lb{5.31}
\Psi[h_{ab}]=C[h_{ab}]\exp\left(\frac{\I}{\hbar}S[h_{ab}]\right)
\ee
and assumes that $C[h_{ab}]$ is a `slowly varying amplitude' and
$S[h_{ab}]$ is a `rapidly varying phase'. This corresponds to
the substitution
\bdm
p^{ab}\longrightarrow \frac{\delta S}{\delta h_{ab}}\ ,
\edm
which is the classical relation for the canonical momentum.
 From (\ref{5.21}) and (\ref{5.22}) one finds then for 
 $S[h_{ab}]$ the equations 
\bea
16\pi G\, G_{abcd}\frac{\delta S}{\delta h_{ab}}\frac{\delta S}{\delta
h_{cd}}-\frac{\sqrt{h}}{16\pi G}(\,{}^{(3)}\!R-2\Lambda) &=& 0\ , \lb{5.32}
\\
D_a\frac{\delta S}{\delta h_{ab}} &=& 0\ . \lb{5.33}
\eea
In the presence of matter one has additional terms. 
The first equation (\ref{5.32}) is the Hamilton-Jacobi equation for
the gravitational field. One can prove that
the four local equations (\ref{5.32}) and (\ref{5.33}) are equivalent
to all ten Einstein equations. 

If non-gravitational fields are present, as we will now assume, a
mixture of this WKB ansatz with the Born-Oppenheimer ansatz from
molecular physics is appropriate \cite{OUP,honnef,SP89}. 
One writes instead of (\ref{5.31}) now
\be
\lb{BO_ansatz}
\Psi[h_{ab},\phi]\equiv\exp\left(\frac{\I}{\hbar}S[h_{ab},\phi]\right),
\ee
where $S[h_{ab},\phi]$ here denotes a complex function that depends on
both the three-metric $h_{ab}$ and the non-gravitational fields
denoted by $\phi$ (usually taken to be a scalar field). 
Plugging this ansatz into the quantum
constraints (\ref{5.21}) and (\ref{5.22}) and performing an expansion
scheme with respect to the square of the Planck mass
$m_{\rm P}=\sqrt{\hbar/G}$, 
\be
\lb{S_expansion}
S[h_{ab},\phi]=m_{\rm P}^2S_0+S_1+m_{\rm P}^{-2}S_2+\ldots,
\ee
one finds at highest order ($m_{\rm
  P}^2$) that $S_0$ depends only on the three-metric $h_{ab}$ and that it
obeys the Hamilton-Jacobi equation (\ref{5.32}) and 
Eq. (\ref{5.33}) for the pure gravitational field. 

The next order ($m_{\rm P}^0$) gives a functional Schr\"odinger
equation for a wave functional $\psi[h_{ab},\phi]$ in the
background spacetime defined from a solution $S_0$ to (\ref{5.32}) and
(\ref{5.33}), where 
\be
\psi[h_{ab},\phi]:=D[h_{ab}]\exp\left(\frac{\I}{\hbar}S_1[h_{ab},\phi]\right),
\ee 
and $D$ obeys the standard WKB prefactor equation (see e.g. Eq. (2.36)
in \cite{honnef}).
This step yields a Tomonaga-Schwinger equation for
$\psi[h_{ab},\phi]$ with respect to a local time functional
$\tau(\X)$ that is defined from the solution $S_0$ by
\be
\lb{tauGR}
\frac{\delta}{\delta\tau(\X)}:=G_{abcd}\frac{\delta
  S_0}{\delta h_{ab}} 
\frac{\delta}{\delta h_{cd}}.
\ee
In spite of its appearance, $\tau$ is not a scalar function
\cite{GK95}. The functional Schr\"odinger equation is obtained by
evaluating $\psi[h_{ab},\phi]$ along a
solution of the classical Einstein equations, $h_{ab}({\X},t)$,
that corresponds to a solution, $S_0[h_{ab}]$, of the Hamilton--Jacobi
equation,
        $\psi[h_{ab}({\X},t),\phi]$.
After a certain choice of lapse and shift functions, $N$ and
$N^a$, has been made, this solution is obtained from
        \be
        \dot{h}_{ab}=NG_{abcd}
        \frac{\delta S_0}{\delta h_{cd}}+
        2D_{(a}N_{b)}\ .               \label{a2.18}
        \ee
Instead of $\psi[h_{ab},\phi]$, we can write
$\vert\psi[h_{ab}]\rangle$ to indicate 
(by the bra-ket notation) that one has a well-defined
(standard) Hilbert space for the non-gravitational field $\phi$.
Defining
        \be
        \frac{\partial}{\partial t}\,|\psi(t)\rangle:=
        \int \D^3 x \,\dot{h}_{ab}({\bf x},t)\,
        \frac{\delta}{\delta h_{ab}({\bf x})}
        |\psi[h_{ab}]\rangle\ ,
        \ee
one finds the functional Schr\"odinger equation for quantized
non-gravitational fields in the chosen external classical gravitational field,
        \begin{eqnarray}
        \I\hbar\frac{\partial}{\partial t}\,
        |\psi(t)\rangle &=& \hat{H}{}^{\rm m}|\psi(t)\rangle\ ,\nonumber \\
        \hat{H}{}^{\rm m} &:=&
        \int \D^3 x \left\{N({\bf x})
        \hat{\mathcal H}{}^{\rm m}_{\perp}({\bf x})+
        N^a({\bf x})\hat{\mathcal H}{}^{\rm m}_a({\bf x})\right\}\ .  
       \label{a2.19}
        \end{eqnarray}
Here, $\hat{H}{}^{\rm m}$ is the non-gravitational Hamiltonian 
in the Schr\"odinger
picture, parametrically depending on the metric
coefficients of the curved spacetime background.
This is the standard approach for obtaining the limit of quantum field
theory in curved spacetime from canonical quantum gravity. Extending
this scheme to higher orders in $m_{\rm P}^2$, one arrives at quantum
gravitational correction terms to this equation \cite{KS91}. These
terms can be used to calculate potentially observable effects such as
corrections to the CMB anisotropy spectrum \cite{BKK16}. 

The Born-Oppenheimer approximation starting from (\ref{BO_ansatz})
provides only part of the understanding why we observe a classical
spacetime. The remaining part is provided by the process of
decoherence. It was suggested in \cite{Zeh86} and elaborated in
\cite{CK87} to using small inhomogeneities such as density
perturbations or tiny gravitational waves as a ``quantum environment''
in configuration space, whose interaction with relevant degrees of
freedom such as the global size of the universe gives rise to their
classical appearance. Technically, this comes from tracing out these
inhomogeneities in the globally entangled quantum states. In
\cite{Paddy89}, Paddy has extended these investigations to more
general situations and found that three-geometries with the same
intrinsic metric but different size contribute decoherently to the
reduced density matrix for the relevant degrees of freedom.
He concludes his paper with the words
\begin{quote}
\ldots the classical nature of the space-time will tend to disappear
as we observe more and more matter modes. Probably, ignorance {\em is}
bliss. 
\end{quote}
The recovery of time in semiclassical gravity raises the question
whether time in quantum gravity can be recovered from a general
solution of the Wheeler-DeWitt solution. The idea was followed
independently by Paddy \cite{Paddy90} and Greensite
\cite{Greensite}. This generalized time is recovered from the phase of
the wave function and used to define a Schr\"odinger-type inner
product where all variables are integrated over {\em except for} this
time. 
A necessary prerequisite for this to work is that the wave function is
complex and that its phase is not a constant. One can then prove that
the first Ehrenfest theorem is valid if this time variable and the
corresponding inner product is used. Unfortunately, only a restricted
class of solutions fulfills all consistency conditions (including the
validity of the second Ehrenfest theorem), so one either has to
abandon this proposal as a solution to the time problem or to use it
as a new type of boundary condition to select physically allowed
solutions \cite{BK96}.

\section{Quantization of conformal (Weyl) gravity}
\label{sec:2}

The role of conformal transformations and of conformal symmetry is of
central interest for gravitational systems at least since Hermann
Weyl's pioneering work from 1918.
Weyl suggested a theory in which not only the direction of a vector
depends on the path along which the vector is transported through spacetime,
but also its length. This means that space distances and time
intervals depend on the path of rods and clocks through spacetime. In
Weyl's theory there exists the freedom to re-scale (``gauge'') rods
and clocks; the metric can be multiplied by an arbitrary positive
spacetime-dependent function,
\begin{equation}
\label{eqn:cftrans}
g_{\mu\nu}(x)\quad\rightarrow\quad
\tilde{g}_{\mu\nu}(x)=\Omega^{2}(x)g_{\mu\nu}(x)\,. 
\end{equation}
This transformation is called {\em conformal transformation} (later
also {\em Weyl transformation}) and is 
an invariance in Weyl's theory. Connected with this freedom is a
new quantity that Weyl identified with the electromagnetic
four-potential, suggesting the idea of a unification between gravity
and electromagnetism.\footnote{For a review and reference to original
  articles, see \cite{Goenner}.} 

Weyl's theory is impressive, but empirically wrong, as soon noticed by
others, in particular Einstein. If it were true, spectral lines, for
example, would depend on the history of the atomic worldlines, because
an atom can be understood as constituting a
clock.\footnote{Recall that the modern time standard is based on the
  hyperfine transitions in caesium-133.} Quite generally, a particle
with rest mass $m$ can be taken as a clock with frequency
\[
\nu=m\frac{c^2}{h},
\]
so a path-dependent frequency would correspond to a path-dependent
rest mass, since $c$ and $h$ are universal units. This is definitely
empirically wrong.

Weyl thus had to give up his
theory, but later used essential elements of his idea to provide the
foundation of modern gauge theory. Einstein, however, was speculating
about the existence of a theory that, while preserving the conformal
invariance of Weyl's theory, 
does not include a hypothesis about the transport of rods and clocks, thus
avoiding the problems of Weyl's theory. In a paper entitled ``\"Uber eine
naheliegende Erg\"anzung des Fundamentes der allgemeinen
Relativit\"atstheorie''
\cite{Einstein21},\footnote{English translation: ``On a Natural
  Addition to the Foundation of the General Theory of Relativity''}
Einstein suggested to use the scalar
\be
\lb{Weyl_scalar}
C_{\mu\nu\lambda\rho} C^{\mu\nu\lambda\rho}
\ee
formed from the Weyl tensor $C_{\mu\nu\lambda\rho}$ as the basis of
this theory.\footnote{In his paper, Einstein acknowledges 
  the help of 
  the Austrian mathematician Wilhelm Wirtinger in his attempt.
In a letter to Einstein sent one day after Einstein's academy talk on
which \cite{Einstein21} is based, Wirtinger suggested as one
possibility to use an action principle based on (\ref{Weyl_scalar}),
see \cite{EinsteinColl}, p.~117.} The Weyl tensor
$C^{\mu}_{\nu\lambda\rho}$
(with one upper component) is invariant under the conformal
transformations (\ref{eqn:cftrans}). 

At the end of his article, Einstein intended to add the following short
summary, which can be found in his hand-written manuscript, but which he deleted
before submission. It reads (our translation from the
German)\footnote{The $\varphi_{\nu}$ denote the components of the
  electromagnetic four-potential.} 
\begin{quotation}
Short summary: it is shown that one can, following Weyl's basic ideas,
develop a theory of invariants on the objective existence of lightcones 
(invariance of the equation $ds^2=0$) alone, which does not, in
contrast to Weyl's 
theory, contain a hypothesis about transport of distances and in which
the potentials $\varphi_{\nu}$ do not enter explicitly the
equations. Later investigations must show whether the theory will be
physically valid.\footnote{The original German reads 
  (\cite{EinsteinColl}, p.~416): ``Kurze 
  Zusammenfassung: Es wird gezeigt, dass man entsprechend dem
  Weyl'schen Grundgedanken auf die objektive Existenz der Lichtkegel
(Invarianz der Gleichung $ds^2=0$) alleine eine Invarianten-theorie
gr\"unden kann, die jedoch im Gegensatz zu Weyl's Theorie keine
Hypothese \"uber Strecken\"ubertragung enth\"alt und in welcher die
Potentiale $\varphi_{\nu}$ nicht explizite in die Gleichungen
eingehen. Ob die Theorie auf physikalische G\"ultigkeit Anspruch
erheben kann, m\"ussen sp\"atere Untersuchungen ergeben.''}
\end{quotation}

In fact, even before Einstein, Rudolf Bach had considered an action
based on (\ref{Weyl_scalar}) and derived and discussed the ensuing
field equations \cite{Bach21}. 
When we talk here of conformal gravity or Weyl gravity, we do not mean
Weyl's original gravitational gauge theory from 1918, but a theory
that is based on the action suggested by Bach, Einstein, and Wirtinger. 
We write the action in the following form:
\begin{equation}
\label{eqn:W-action}
S^{\ssst \rm W}:=-\frac{\alpha_{\ssst\rm W}\hbar}{4}\int{\rm
  d}^4x\sqrt{-g}\,C_{\mu\nu\lambda\rho} 
C^{\mu\nu\lambda\rho}\,,
\end{equation}
where $\alpha_{\ssst\rm W}$ is a dimensionless coupling constant.
We have introduced Planck's constant $\hbar$ (which, of course, is
irrelevant for the classical theory) for two reasons: first, it gives
the correct dimensions for the action and renders the 
constant $\alpha_{\ssst\rm W}$ dimensionless and, second, $S^{\ssst
  \rm W}/\hbar$ is the relevant quantity in the quantum theory on
which we will focus in our paper; in fact, this will suggest the
semiclassical expansion scheme with respect to  $\alpha_{\ssst\rm W}$
presented below.

The theory based on (\ref{eqn:W-action}) was discussed at both the
classical and quantum level \cite{Mannheim12}. At the classical level,
it was used, for example, to explain galactic rotation curves without
the need for dark matter, although this explanation has met severe
criticism \cite{PX95}. At the quantum level, it is a candidate for a
renormalizable theory of quantum gravity, although it seems to violate
unitarity.\footnote{We write ``seems'', because the ghosts connected
  with non-unitarity may be removable \cite{Maldacena11}.} We adopt
here the point of view to take (\ref{eqn:W-action}) as the starting
point for a conformally invariant gravity theory, for which a
semiclassical expansion scheme can be applied to its canonically
quantized version and compared with the scheme for  
quantized GR.
 We do not assume that (\ref{eqn:W-action}) is a candidate for an
 alternative to GR. In the following, we shall study
 the canonical structure of this theory. Our treatment is based on the
 more general treatment presented in \cite{KN16a}.

In order to deal with higher-derivative theories such as
(\ref{eqn:W-action}),\footnote{For the history of such theories, see
  for example \cite{Schmidt06}.} it is convenient to reduce the order by
introducing new independent variables. In our case, this is achieved
by introducing the extrinsic curvature $K_{ij}$, which in general
relativity is a function of the time derivative of the three-metric 
$h_{ij}$. This can be implemented in the canonical formalism by adding 
a constraint $\lambda^{ij}\left(2K_{ij}-\mathcal{L}_{n}h_{ij}\right)$,
where $\lambda^{ij}$ is a Lagrange multiplier. There are also other
methods to ``hide'' the second derivative of the three-metric
\cite{Boulware83,DerrHD}. 
 
In order to manifestly reveal conformal invariance of the Weyl action,
we will use here an irreducible decomposition of the 3-metric into its
scale part\footnote{In quantum GR, there exist attempts to
  quantize solely the conformal factor \cite{Paddy85}. Paddy has
  derived from this the interesting conclusion that the Planck length
  provides a lower bound to measurable physical lengths. The situation
  will be different here, because Weyl gravity does not contain an
  intrinsic length scale.} 
\be
\label{a_definition}
a={(\sqrt{h})}^{\ssst 1/3}
\ee
and its conformally invariant (unimodular) part
\be 
\label{hbar_definition}
\bar{h}_{ij}=a^{-2}h_{ij}.
\ee
In addition, we define the following variables \cite{KN16a}
\bea
\label{eqn:VarsN-m}
&\bar{N}^{i}=N^{i}\,,\quad\bar{N}_{i}=a^{-2} N_{i}\,,\quad\bar{N}=a^{-1}N\,,\\
\label{eqn:VarsKbar}
& \bar{K}_{ij}^{\ssst\rm
  T}=a^{-1}{K}_{ij}^{\ssst\rm
  T}\,,\quad\,\,\,
\bar{K}=\frac{aK}{3}\,, 
\eea
where $\bar{K}_{ij}^{\ssst\rm T}$ and $\bar{K}$ are the rescaled
traceless and trace parts of the extrinsic curvature, respectively;
as a consequence of the decomposition of the three-metric
(\ref{a_definition}) and (\ref{hbar_definition})
they are given by 
\begin{equation}
\label{eqn:K-split}
\bar{K}_{ij}^{\ssst\rm
  T}=\frac{1}{\bar{N}}\left(\dot{\bar{h}}_{ij}-2\left[D_{(i}\bar{N}_{j)}\right]^{\ssst\rm  
    T}\right)\,,\quad
\bar{K}=\frac{1}{\bar{N}}\left(\frac{\dot{a}}{a}-\frac{1}{3}D_{i}N^{i}\right)\,. 
\end{equation}
It can be shown by direct calculation that
$\left[D_{(i}\bar{N}_{j)}\right]^{\ssst\rm T}$ is independent of $a$
and that $\bar{K}_{ij}^{\ssst\rm T}$ is the conformally invariant part of the
extrinsic curvature. We refer to the variables in
(\ref{eqn:VarsN-m}) and (\ref{eqn:VarsKbar})
as \textit{unimodular-conformal variables}, and we will formulate the
canonical theory in terms of them. The advantage of using these
variables is that only the scale $a$ and the trace $\bar{K}$ transform
under conformal transformation, 
\begin{equation}
\label{eqn:K-transf}
\bar{K} \quad \rightarrow \quad\bar{K} + \bar{n}^{\mu}\partial_{\mu}\log \Omega\,,\qquad a \quad\rightarrow \quad \Omega a \,,
\end{equation}
where $\bar{n}^{\mu}=a n^{\mu}$ and
$\bar{n}_{\mu}=a^{-1}{n}_{\mu}$. This significantly simplifies
the canonical formulation and makes conformal invariance of the theory
manifest, since the only two variables affected by conformal
transformation completely vanish from the constraints, as will be
shown below. 

The canonical approach employs a $3+1$ decomposition of spacetime
quantities. For GR, this is the standard ADM approach \cite{OUP}. In
the present case, one has to perform a $3+1$ decomposition of the Weyl
tensor, which can be found, for example, in \cite{ILP,KOT14}.
The constrained 3+1-decomposed Lagrangian density of the Weyl action
in terms of the unimodular-conformal variables introduced above then becomes 
\begin{eqnarray}
\label{eqn:LcW}
\mathcal{L}_{c}^{\ssst\rm
  W}&=&\bar{N}\Biggl\lbrace-\frac{\alpha_{\ssst\rm W}\hbar}{2}
\bar{h}^{ia}\bar{h}^{jb}\bar{C}_{ij}^{\ssst\rm
  T}\bar{C}_{ab}^{\ssst\rm T}+\alpha_{\ssst\rm W}\hbar
\bar{C}_{ijk}^{2}-a^{5}\lambda^{ij\ssst\rm
  T}\left[2\bar{K}_{ij}^{\ssst\rm
    T}-\frac{1}{\bar{N}}\left(\dot{\bar{h}}_{ij}-2\left[D_{(i}\bar{N}_{j)}\right]^{\ssst\rm 
      T}\right)\right]\nonumber\\ 
\;\; & & -2a^3\lambda
\left[\bar{K}-\frac{1}{\bar{N}}\left(\frac{\dot{a}}{a}-\frac{1}{3}D_{a}N^{a}\right)\right]\Biggr\rbrace\,, 
\end{eqnarray}
where $\lambda^{ij\ssst\rm T}$ and $\lambda$ are traceless and trace
parts of the Lagrange multiplier $\lambda^{ij}$, and 
\begin{equation}
\label{eqn:elWumod}
\bar{C}_{ij}^{\ssst\rm T}=\mathcal{L}_{\bar{n}}\bar{K}_{ij}^{\ssst\rm T}-\frac{2}{3}\bar{h}_{ij}\bar{K}_{ab}^{\ssst \rm T}\bar{h}^{an}\bar{h}^{bm}\bar{K}_{nm}^{\ssst \rm T}-\!\,^{\ssst (3)}\!R_{ij}^{\ssst\rm T}-\frac{1}{\bar{N}}\left[ D_{i}D_{j}\right]^{\ssst \rm T}\bar{N}\,
\end{equation}
is the ``electric part'' of the Weyl tensor, containing only
velocities of the traceless part of the extrinsic curvature, and 
\begin{equation}
\label{eqn:magWumod}
\bar{C}_{ijk}=2\delta_{i}^{[d}\left(\delta_{j}^{e]}\delta_{k}^{f}-\bar{h}_{jk}\bar{h}^{e]f}\right)D_{d}\bar{K}_{ef}\,,\quad\bar{C}_{ijk}^2\equiv\bar{C}_{ijk}\bar{h}^{ia}\bar{h}^{jb}\bar{h}^{kc}\bar{C}_{abc}  
\end{equation}
is related to the ``magnetic part'' of the Weyl tensor, as explained
in \cite{ILP}. The second expression in (\ref{eqn:magWumod}) should not
contain any traces $\bar{K}$ and therefore be conformally invariant,
but we assume this without proof. Each object with superscript
``$\ssst\rm T$'' is traceless. It
can be shown easily that the trace of the sum of the first two terms
in (\ref{eqn:elWumod}) vanishes, that is, that
$h^{ij}\mathcal{L}_{n}\bar{K}^{\ssst\rm
  T}_{ij}=2a^{-2}\bar{K}_{ab}^{\ssst \rm
  T}\bar{h}^{an}\bar{h}^{bm}\bar{K}_{nm}^{\ssst \rm T}$. 

We now take unimodular-conformal variables (\ref{a_definition}),
(\ref{hbar_definition}, 
(\ref{eqn:VarsN-m}), and (\ref{eqn:VarsKbar}) and 
derive their conjugate momenta in the standard way,
\begin{eqnarray}
\label{eqn:p}
& p_{\ssst \bar{N}}=\frac{\del \mathcal{L}^{\ssst\rm W}_{c}}{\del
  \dot{\bar{N}}}\approx 0\,,\quad p^{i}=\frac{\del
  \mathcal{L}^{\ssst\rm W}_{c}}{\del \dot{N}^{i}}\approx
0\,,\quad\bar{P}=\frac{\del \mathcal{L}^{\ssst\rm W}_{c}}{\del
  \dot{\bar{K}}}\approx 0\,,\\ 
&\quad\bar{p}^{ij}=\frac{\del \mathcal{L}^{\ssst\rm W}_{c}}{\del
  \dot{\bar{h}}_{ij}}=a^{5}\lambda^{ij \ssst\rm T}\,,\quad 
p_{a}=\frac{\del \mathcal{L}^{\ssst\rm W}_{c}}{\del \dot{a}}=2a^2 \lambda\,,\\
&\quad\bar{P}^{ij}=\frac{\del \mathcal{L}^{\ssst\rm W}_{c}}{\del
  \dot{\bar{K}}_{ij}^{\ssst\rm T}}=-\alpha_{\ssst\rm
  W}\hbar\,\bar{h}^{ia}\bar{h}^{jb}\bar{C}_{ab}^{\ssst\rm T}\,.  
\end{eqnarray}
Note that the momenta $\bar{p}^{ij}$ and $\bar{P}^{ij}$ are traceless. The
novelty with respect to GR is the emergence of another primary constraint,
$\bar{P}\approx 0$; this suggests that $\bar{K}$ is arbitrary, in the
same manner as $p_{\bar{N}}\approx 0$ and $p_{i}\approx 0$ suggest
that $\bar{N}$ and $N^{i}$ are arbitrary. 

It can easily be checked
that the transformation from the original variables to the
unimodular-conformal variables is a canonical one. The
Poisson brackets (PBs) of the variables are 
\begin{equation}
\label{eqn:PBvars1}
\left\lbrace q_{ij}^{A}(\vec{x}),\Pi^{ab}_{B}(\vec{y})
\right\rbrace=\left(\delta^{a}_{(i}\delta^{b}_{j)}-\frac{1}{3}h_{ij}h^{ab}\right)\delta_{B}^{A}\delta 
(\vec{x},\vec{y})\,,\quad \left\lbrace q^{A}(\vec{x}),\Pi_{B}(\vec{y})
\right\rbrace=\delta_{B}^{A}\delta (\vec{x},\vec{y})\,, 
\end{equation}
where $q_{ij}^{A}=( \bar{h}_{ij},\bar{K}_{ij}^{\ssst\rm
  T}),\,\Pi^{ab}_{B}=( \bar{p}^{ab},\bar{P}^{ab})$ are the variables
in the conformally
invariant subspace of phase space, and $q^{A}=(a, \bar{K})\,,\Pi_{B}=(p_{a},
\bar{P})$ is the scale-trace subspace of phase space (and similar
for the lapse-shift sector). All other PBs vanish. 

After performing the Legendre transformation (from which
$\dot{\bar{K}}\bar{P}$ is absent, since $\dot{\bar{K}}$ does not
appear in the Lagrangian) and investigating the emerging constraints,
we can write the total Hamiltonian as 
\begin{equation}
\label{eqn:WHam}
H^{\ssst\rm W}=\intx\Biggl\lbrace \bar{N}\mathcal{H}^{\ssst\rm
  W}_{\bot}+N^{i}\mathcal{H}^{\ssst\rm
  W}_{i}+\left(\bar{N}\bar{K}-\frac{1}{3}D_{i}
  N^{i}\right)\mathcal{Q}^{\ssst\rm
  W}+\lambda_{\ssst\bar{N}}p_{\ssst\bar{N}}+\lambda_{i}p^{i}+\lambda_{\ssst\bar{P}}\bar{P}\Biggr\rbrace+H_{\rm
  surf}\,, 
\end{equation}
from which one finds the secondary constraints
\begin{eqnarray}
\label{eqn:HamWcf}
&\mathcal{H}^{\ssst\rm
  W}_{\bot}=-\frac{\bar{h}_{ik}\bar{h}_{jl}\bar{P}^{ij}\bar{P}^{kl}}{2\alpha_{\ssst\rm
    W}\hbar}+\left(\,^{\ssst (3)}\! R_{ij}^{\ssst\rm
    T}+D_{i}D_{j}\right)\bar{P}^{ij}+2\bar{K}_{ij}^{\ssst\rm
  T}\bar{p}^{ij}-\alpha_{\ssst\rm W}\hbar\bar{C}_{ijk}^2\approx 0\,,\\ 
\label{eqn:MomWcf}
&\mathcal{H}^{\ssst\rm W}_{i}=-2\del_{k}\left(
\bar{h}_{ij}\bar{p}^{jk}\right)+\del_{i}\bar{h}_{jk}\bar{p}^{jk}-2\del_{k}\left(\bar{K}_{ij}^{\ssst\rm
  T}\bar{P}^{jk}\right)+\del_{i}\bar{K}_{jk}^{\ssst\rm
T}\bar{P}^{jk}\approx 0\,,\\ 
\label{eqn:Qw}
&\mathcal{Q}^{\ssst\rm W}=ap_{a}\approx 0\,.
\end{eqnarray}
The first two are the Hamiltonian and momentum constraints, and they
are analogous to (\ref{5.21}) and (\ref{5.22}), although the
structure of the Hamiltonian constraint is significantly
different. The new constraint (\ref{eqn:Qw}) comes from the consistency
condition for the primary constraint $\bar{P}\approx 0$, 
\begin{equation}
\label{eqn:Pdot}
\dot{\bar{P}}=\left\lbrace\bar{P},H^{\ssst\rm
    W}\right\rbrace=-\frac{\del H^{\ssst\rm W}}{\del
  \bar{K}}=-\bar{N}a\, p_{a}\stackrel{!}{\approx} 0\,. 
\end{equation}
A brief inspection of constraints reveals that the Hamiltonian and
momentum constraints are manifestly conformally invariant, due to
the use of the unimodular-conformal variables\footnote{It can be shown
  that terms in $\left(\,^{\ssst (3)}\! R_{ij}^{\ssst\rm
      T}+D_{i}D_{j}\right)\bar{P}^{ij}$ which depend on $a$ cancel,
  making this expression conformally invariant.} --- the Hamiltonian
and momentum constraints are independent of the scale $a$ and trace
$\bar{K}$. The constraints $\bar{P}$ and $\mathcal{Q}^{\ssst\rm W}$
commute, and they also commute with the rest of the constraints. The
Hamiltonian and the momentum constraints close the same hypersurface
foliation algebra as in GR, see \cite{DerrHD}. This is expected for
any reparametrization invariant metric theory, see \cite{Thiemann}, 
p.~57. Hence, all constraints are first class. 

Therefore, the Hamiltonian and momentum constraints have the same
meaning as in GR. The momentum constraint is extended to include the
extrinsic curvature sector, since the components of $K_{ij}$ are treated
as independent variables in this higher-derivative theory. Thus the
three-dimensional diffeomorphism invariance now includes changes of
$\bar{K}_{ij}^{\ssst\rm T}$. But what is the meaning of the $\bar{P}$ and
$\mathcal{Q}^{\ssst\rm W}$ constraints? It can be shown that these
constraints comprise a generator of conformal \textit{gauge}
transformation, as shown in \cite{ILP} in terms of the original
variables (which also include the lapse, prone to conformal
transformation). In unimodular-conformal variables, a procedure
similar to \cite{ILP} leads to the following generator of conformal
transformation \cite{KN16a}: 
\begin{equation}
\label{eqn:confG}
G_{\omega}^{\ssst\rm W}[\omega,\dot{\omega}]=\intx
\left(\,\mathcal{Q}^{\ssst\rm
    W}\omega+\bar{P}\mathcal{L}_{\bar{n}}\omega\right)=\intx \left(a
  p_{a}\omega+\bar{P}\mathcal{L}_{\bar{n}}\omega\right)\,, 
\end{equation}
which generates here a transformation only for the scale $a$ and the trace $\bar{K}$.
We emphasize that primary and secondary constraints have to appear together to ensure a correct transformation,
as emphasized in particular by Pitts \cite{Pitts14}. 

 A closer look at the Hamiltonian constraint (\ref{eqn:HamWcf})
reveals that the ``intrinsic time'' of GR contained in the scale part
$a$ is absent. This is not surprising, because we are dealing here
with a conformally invariant theory. The ``problem of time'' in
quantum gravity \cite{OUP} is for the Weyl theory thus of a different
nature than for GR. This difference will also be relevant for the recovery of
semiclassical time discussed below.
 
Let us now turn to configuration space. In analogy to the Hamilton-Jacobi
function of GR, Eq. (\ref{5.32}), one can define a Hamilton-Jacobi
functional in Weyl gravity as well, which is defined on
full configuration space, 
\[ 
S^{\ssst\rm W}=S^{\ssst\rm W}[\bar{h}_{ij},a,\bar{K}_{ij}^{\ssst\rm T},\bar{K}].
\]
The conjugate momenta $\bar{p}^{ij}$ and $\bar{P}^{ij}$ follow form
this functional in the usual way,
\begin{equation}
\label{eqn:HJmomC}
\bar{p}^{ij}=\frac{\delta S^{\ssst\rm W}}{\delta
  \bar{h}_{ij}}\,,\quad\bar{P}^{ij}=\frac{\delta S^{\ssst\rm
    W}}{\delta \bar{K}_{ij}^{\ssst\rm T}}\,,\quad p_{a}=\frac{\delta
  S^{\ssst\rm W}}{\delta a}\,,\quad\bar{P}=\frac{\delta S^{\ssst\rm
    W}}{\delta \bar{K}}\,. 
\end{equation} 
Due to the primary-secondary pair of constraints $\bar{P}\approx 0$
and $\mathcal{Q}^{\ssst\rm W}\approx 0$, we can conclude, however, that
the functional $S^{\ssst\rm W}$ does not depend on $a$ and $\bar{K}$,
since its infinitesimal conformal variations vanish, 
\begin{equation}
\label{eqn:aktg}
\frac{\delta S^{\ssst\rm W}}{\delta a}=0\,,\quad\frac{\delta
  S^{\ssst\rm W}}{\delta
  \bar{K}}=0\quad\Rightarrow\quad\delta_{\omega}S^{\ssst\rm
  W}=\intx\left(\frac{\delta S^{\ssst\rm W}}{\delta a}\,\delta
  a+\frac{\delta S^{\ssst\rm W}}{\delta \bar{K}}\,\delta
  \bar{K}\right)=0\,. 
\end{equation}
One can then interpret $S^{\ssst\rm W}$ as a conformally invariant
functional solving the conformally invariant
\textit{Weyl-Hamilton-Jacobi equation} (WHJ equation)
obtained from (\ref{eqn:HamWcf}), 
\begin{equation} 
\label{eqn:WHJ}
-\frac{1}{2\alpha_{\ssst\rm
    W}\hbar}\bar{h}_{ik}\bar{h}_{jl}\frac{\delta S^{\ssst\rm
    W}}{\delta \bar{K}_{ij}^{\ssst\rm T}}\frac{\delta S^{\ssst\rm
    W}}{\delta \bar{K}_{kl}^{\ssst\rm T}}+\left(\,^{\ssst (3)}\!
  R_{ij}^{\ssst\rm T}+D_{i}D_{j}\right)\frac{\delta S^{\ssst\rm
    W}}{\delta \bar{K}_{ij}^{\ssst\rm T}}+2\bar{K}_{ij}^{\ssst\rm
  T}\frac{\delta S^{\ssst\rm W}}{\delta \bar{h}_{ij}}-\alpha_{\ssst\rm
  W}\hbar\bar{C}_{ijk}^2= 0\,. 
\end{equation}
We expect that $S^{\ssst\rm W}$, as a solution to the
above equation, gives a ``classical trajectory'' in the
configuration subspace spanned by $\left\lbrace
  \bar{h}_{ij},\bar{K}_{ij}^{\ssst\rm T}\right\rbrace$. Due to
(\ref{eqn:aktg}), a tangent to this trajectory does not have components
in the $a$ and $\bar{K}$ directions of the configuration space. In other
words, the classical state of this theory does not follow directions
along changes of $a$ and $\bar{K}$ in configuration space.  

Quantization is now performed in the sense of Dirac by implementing
the classical constraints as restrictions on physically allowed wave
functionals on the full ocnfiguration space \cite{OUP},
\[
\Psi\equiv\Psi[\bar{h}_{ij},a,\bar{K}_{ij}^{\ssst\rm T},\bar{K}].
\]
The canonical variables are promoted into operators in the standard
way, 
\begin{eqnarray}
&\hat{\bar{h}}_{ij}(x)\Psi=
\bar{h}_{ij}(x)\Psi\,,\quad\hat{\bar{p}}^{ij}(x)\Psi=
-i\hbar\frac{\delta }{\delta \bar{h}_{ij}(x)}\Psi\,,\\ 
&\hat{\bar{K}}_{ij}^{\ssst\rm T}(x)\Psi= \bar{K}_{ij}^{\ssst\rm
  T}(x)\Psi\,,\quad\hat{\bar{P}}^{ij}(x)\Psi= -{\rm i}\hbar\frac{\delta
}{\delta \bar{K}_{ij}^{\ssst\rm T}(x)}\Psi\,,\\ 
&\hat{a}(x)\Psi= a(x)\Psi\,,\quad \hat{\bar{p}}_{a}(x)\Psi=
-{\rm i}\hbar\frac{\delta }{\delta a(x)}\Psi\,,\\ 
&\hat{\bar{K}}(x)\Psi= \bar{K}(x)\Psi\,,\quad \hat{\bar{P}}(x)\Psi=
-{\rm i}\hbar\frac{\delta }{\delta \bar{K}(x)}\Psi. 
\end{eqnarray}
The quantization of the constraints yields \cite{KN16b}
\begin{equation}
\hat{\mathcal{H}}^{\ssst\rm W}_{\bot}\Psi=0\,,\quad \hat{\mathcal{H}}^{\ssst\rm W}_{i}\Psi=0\,,\quad \hat{\bar{P}}\Psi=0\,,\quad \hat{\mathcal{Q}}^{\ssst\rm W}\Psi =0.
\end{equation}  
 
The first of these equations is the quantized Hamiltonian constraint,
which replaces the WDW equation of quantum GR and which we will
therefore call the  ``Weyl-Wheeler-DeWitt'' (WWDW) equation.
Neglecting here the ubiquitous factor ordering problem, it assumes the
explicit form
\begin{eqnarray}
\label{eqn:W-WDW}
& & \Biggl[\frac{\hbar}{2\alpha_{\ssst\rm
    W}}\bar{h}_{ik}\bar{h}_{jl}\frac{\delta^2}{\delta
  \bar{K}_{ij}^{\ssst\rm T}\delta \bar{K}_{kl}^{\ssst\rm
    T}}-{\rm i}\hbar\left(\,^{\ssst (3)}\! R_{ij}^{\ssst\rm
    T}+D_{ij}^{\ssst\rm T}\right)\frac{\delta}{\delta
  \bar{K}_{ij}^{\ssst\rm T}}-2{\rm i}\hbar \bar{K}_{ij}^{\ssst\rm
  T}\frac{\delta}{\delta \bar{h}_{ij}} \nonumber \\
  & & \ \  -\alpha_{\ssst\rm
  W}\hbar\bar{C}_{ijk\bot}^2+\hat{\mathcal{H}}^{\rm m}_{\perp}\Biggl]\Psi=0\,. 
\end{eqnarray}
One recognizes that the WWDW equation is structurally different from
the WDW equation, since the wave functional does not depend only on
the three-metric, but also on its evolution (the second fundamental
form). There is also no scale $a$ present and therefore no intrinsic time
in the sense of the WDW equation; there is no indefinite ``DeWitt
metric''. It is also interesting to see that $\hbar$ drops out 
after dividing the whole equation by $\hbar$.
 Formally this is due to our use of $\alpha_{\rm W}\hbar$ in
the action instead of just $\alpha_{\rm W}$; re-scaling $\alpha_{\rm
  W}\to \alpha_{\rm W}/\hbar$ would bring back $\hbar$ at the places
similar to the Wheeler-DeWitt equation (\ref{5.21}), but the important
point is that $\hbar$ can be made to disappear by a simple
re-scaling. 
This is, of course, a property of the vacuum theory. If we add a
matter Hamiltonian density to the WWDW equation, as we shall do below,
$\hbar$ will not disappear when dividing the whole equation by
$\hbar$. 

The quantum momentum constraints read
\begin{eqnarray}
\label{eqn:W-Qmom}
& & \hat{\mathcal{H}}^{\ssst\rm W}_{i}\Psi={\rm i}\hbar\Biggl[2\del_{k}\left(
\bar{h}_{ij}\frac{\delta \Psi}{\delta
  \bar{h}_{jk}}\right)-\del_{i}\bar{h}_{jk}\frac{\delta }{\delta
\Psi\bar{h}_{jk}}+2\del_{k}\left(\bar{K}_{ij}^{\ssst\rm T}\frac{\delta
  \Psi }{\delta \bar{K}_{jk}^{\ssst\rm
    T}(x)}\right)\nonumber\\
 & & \ \    -\del_{i}\bar{K}_{jk}^{\ssst\rm T}
 +\frac{\delta \Psi
}{\delta \bar{K}_{jk}^{\ssst\rm T}}+\hat{\mathcal{H}}^{\rm m}_{i}\Psi\Biggr]= 0 ,
\end{eqnarray}
or alternatively, in a manifestly covariant version,
\begin{equation}
\label{eqn:W-QmomC}
\hat{\mathcal{H}}^{\ssst\rm W}_{i}\Psi={\rm i}\hbar\Biggl[2D_{k}\left(
\bar{h}_{ij}\frac{\delta \Psi}{\delta
  \bar{h}_{jk}}\right)+2D_{k}\left(\bar{K}_{ij}^{\ssst\rm
  T}\frac{\delta \Psi }{\delta \bar{K}_{jk}^{\ssst\rm
    T}}\right)-D_{i}\bar{K}_{jk}^{\ssst\rm T}\frac{\delta \Psi
}{\delta \bar{K}_{jk}^{\ssst\rm T}}+\hat{\mathcal{H}}^{\rm m}_{i}\Psi\Biggr]= 0\,. 
\end{equation}
Finally, the new quantum constraints read
\begin{equation}
\label{eqn:W-confQ}
\frac{\delta \Psi }{\delta \bar{K}}=0\,,\quad a\frac{\delta \Psi}{\delta a}=0\,.
\end{equation}  
The meaning of (\ref{eqn:W-confQ}) is obvious: the wave functional
does not depend on $a$ and $\bar{K}$;hence, it is conformally
invariant (apart from a possible phase factor). This is a direct
consequence of the first class nature of the
constraints $\bar{P}=0$ and $\mathcal{Q}^{\ssst\rm W}=ap_{a}=0$. Thus,
we have a conformally invariant canonical quantum gravity theory derived
from the Weyl action. Equivalently, one could have started from a
reduced phase space without $a$ and $K$ and ended up with (\ref{eqn:W-WDW}) and
(\ref{eqn:W-Qmom}) only, with $\Psi$ depending on 10 (instead of 12)
configuration variables from the start. 

Looking at the whole picture, we conclude that solutions to the
WWDW equation are conformally invariant (scale and trace
independent), and are indistinguishable for two three-metrics that are
conformal to each other.

\section{Semiclassical Weyl gravity and the recovery of time}
\label{sec:3}

We consider quantum Weyl gravity with a conformally coupled matter
field $\phi$, for conformal matter does not spoil the first-class 
nature of constraints; it only modifies their explicit form. We can then
quantize the theory while preserving its conformal invariance. 

In the spirit of the semiclassical (Born-Oppenheimer type) expansion
for the WDW equation, 
we make an ansatz for the wave functional in which the
``heavy'' part, being the pure gravitational part, is separated from
the matter part \cite{KN16b}. We write for the full quantum state
in analogy to (\ref{BO_ansatz})
\begin{equation}
\label{eqn:W-sclas}
\Psi\left[\bar{h}_{ij},\bar{K}_{ij}^{\ssst\rm T},\phi\right]
\equiv
\exp\left(\frac{\I}{\hbar}S\left[\bar{h}_{ij},\bar{K}_{ij}^{\ssst\rm
      T},\phi\right]\right). 
\end{equation}
Plugging (\ref{eqn:W-sclas}) into the WWDW equation
(\ref{eqn:W-WDW}) gives
\begin{eqnarray}
\label{eq-Wsemiclmaster}
\frac{{\rm i}}{2\alpha_{\ssst\rm W}}\bar{h}_{ik}\bar{h}_{jl}\frac{\delta^2
  S}{\delta \bar{K}_{ij}^{\ssst\rm T}\delta \bar{K}_{kl}^{\ssst\rm
    T}}-\frac{1}{2\alpha_{\ssst\rm
    W}\hbar}\bar{h}_{ik}\bar{h}_{jl}\frac{\delta S}{\delta
  \bar{K}_{ij}^{\ssst\rm T}}\frac{\delta S}{\delta
  \bar{K}_{kl}^{\ssst\rm T}}+\left(\,^{\ssst (3)}\! R_{ij}^{\ssst\rm
    T}+D_{ij}^{\ssst\rm T}\right)\frac{\delta S}{\delta
  \bar{K}_{ij}^{\ssst\rm T}}\nonumber\\ 
+2\bar{K}_{ij}^{\ssst\rm T}\frac{\delta S}{\delta 
  \bar{h}_{ij}}-\alpha_{\ssst\rm
  W}\hbar\bar{C}_{ijk}^2+\frac{\left(\hat{\mathcal{H}}^{\rm m}_{\bot}\Psi\right)}{\Psi}=0. 
\end{eqnarray} 
The expansion can be performed with respect to $\alpha^{-1}_{\ssst\rm
  W}$, for this 
coupling constant appears at the same place (in the kinetic
term and in part of the potential) as $m_{\rm P}^{2}$ appears in the WDW
equation. The functional $S$ can then be expanded in powers of
$\alpha^{-1}_{\ssst\rm W}\ll 1$, assuming $\alpha_{\ssst\rm W}$ to be
large; this is similar to the Planck-mass expansion for quantum GR,
see (\ref{S_expansion}) above. 
 Note that $\alpha_{\ssst\rm W}$ is a dimensionless quantity,
unlike the Planck mass in the case of the WDW equation. This is similar to the
semiclassical expansion of quantum electrodynamics, with the (dimensionless)
fine structure constant as the appropriate expansion parameter \cite{KPS91}. 
We thus write
\begin{equation}
\label{eqn:W-WKB}
S=\alpha_{\ssst\rm W}\sum_{n=0}^{\infty}\left( \frac{1}{\alpha_{\ssst\rm W}}\right)^{n}S^{\ssst\rm W}_{n}\,.
\end{equation}
Note that $\left(\hat{\mathcal{H}}^{\rm m}_{\bot}\Psi\right)/\Psi$, when
expanded in powers of $\alpha_{\ssst\rm W}$, is at most of the order $\alpha_{\ssst\rm W}^2$,
since the highest derivative with respect to the matter field $\phi$
in $\bar{\mathcal{H}}^{\rm m}_{\bot}$ is the second order, which is the
kinetic term (we assume it is the only one of that kind). We shall
then denote with
$\left(\left(\hat{\mathcal{H}}^{\rm m}_{\bot}\Psi\right)/\Psi\right)^{(n)}$,
$n\leq 2$, terms proportional to $\alpha_{\ssst\rm W}^n$. 

Inserting the ansatz (\ref{eqn:W-WKB}) 
 into the WWDW equation and collecting the powers of $\alpha_{\ssst\rm
   W}^2$, we find 
\begin{equation}
\label{eqn:W-sc2}
\alpha_{\ssst\rm W}^{2}: \hspace{10pt}
\left(\left(\hat{\mathcal{H}}^{\rm m}_{\bot}\Psi\right)/\Psi\right)^{(2)}=0\quad\Rightarrow\quad
\frac{\delta S^{\ssst\rm W}_{0}}{\delta \phi}=0. 
\end{equation}
This is analogous to the situation in GR \cite{KS91}. At the next
order, $\alpha_{\ssst\rm W}$, we have 
\begin{eqnarray}
\label{eqn:W-sc1}
\alpha_{\ssst\rm W}^{1}:\quad -\frac{1}{2\hbar}\bar{h}_{ik}\bar{h}_{jl}\frac{\delta
  S^{\ssst\rm W}_0}{\delta \bar{K}_{ij}^{\ssst\rm T}}\frac{\delta
  S^{\ssst\rm W}_0}{\delta \bar{K}_{kl}^{\ssst\rm T}}+\left(\,^{\ssst
    (3)}\! R_{ij}^{\ssst\rm T}+D_{ij}^{\ssst\rm T}\right)\frac{\delta
  S^{\ssst\rm W}_0}{\delta \bar{K}_{ij}^{\ssst\rm
    T}}+2\bar{K}_{ij}^{\ssst\rm T}\frac{\delta S^{\ssst\rm
    W}_0}{\delta \bar{h}_{ij}}-\hbar\bar{C}_{ijk}^2= 0\,, 
\end{eqnarray}
which is nothing else than the Weyl-HJ equation
(\ref{eqn:WHJ}), with $S^{\ssst\rm W}\equiv \alpha_{\ssst\rm
  W}S^{\ssst\rm W}_0$. 

At the next order, ($\alpha_{\ssst\rm W}^0$), we obtain
\begin{eqnarray}
\label{eqn:W-sc0}
\alpha_{\ssst\rm W}^{0}: \quad \frac{{\rm
    i}}{2}\bar{h}_{ik}\bar{h}_{jl}\frac{\delta^2 
  S^{\ssst\rm W}_{0}}{\delta \bar{K}_{ij}\delta
  \bar{K}_{kl}}-\frac{1}{2\hbar}\bar{h}_{ik}\bar{h}_{jl}\frac{\delta
  S^{\ssst\rm W}_0}{\delta \bar{K}_{ij}^{\ssst\rm T}}\frac{\delta
  S^{\ssst\rm W}_1}{\delta \bar{K}_{kl}^{\ssst\rm T}}+\left(\,^{\ssst
    (3)}\! R_{ij}^{\ssst\rm T}+D_{ij}^{\ssst\rm T}\right)\frac{\delta
  S^{\ssst\rm W}_1}{\delta \bar{K}_{ij}^{\ssst\rm T}}\nonumber\\ 
+2\bar{K}_{ij}\frac{\delta S^{\ssst\rm W}_{1}}{\delta \bar{h}_{ij}}+\left(\left(\hat{\mathcal{H}}^{\rm m}_{\bot}\Psi\right)/\Psi\right)^{(0)}=0\,.
\end{eqnarray}
A procedure analogous to the one used to arrive at the functional Schr\"odinger equation in quantum GR motivates us to propose the following functional:
\begin{equation}
f\equiv D[\bar{h}_{ij},\bar{K}_{ij}^{\ssst\rm T}] \exp\left(\frac{\I}{\hbar}S^{\ssst\rm W}_{1}\right),
\end{equation}
with a condition on the ``WKB prefactor''
$D$ that will be derived below. We first calculate the
following functional derivatives: 
\begin{eqnarray*}
& & {\rm i} \bar{h}_{ik}\bar{h}_{jl}\frac{\delta S^{\ssst\rm W}_{0}}{\delta
  \bar{K}_{ij}^{\ssst\rm T}}\frac{\delta f}{\delta
  \bar{K}_{kl}}={\rm i}\bar{h}_{ik}\bar{h}_{jl}\frac{\delta S^{\ssst\rm
    W}_{0}}{\delta \bar{K}_{ij}^{\ssst\rm T}}\frac{\delta D}{\delta
  \bar{K}_{kl}^{\ssst\rm
    T}}\frac{1}{D}f-\frac{1}{\hbar}\bar{h}_{ik}\bar{h}_{jl}\frac{\delta
  S^{\ssst\rm W}_{0}}{\delta \bar{K}_{ij}^{\ssst\rm T}}\frac{\delta
  S^{\ssst\rm W}_{1}}{\delta \bar{K}_{kl}^{\ssst\rm T}}f,\\
& & 
-2{\rm i}\hbar \bar{K}_{ij}^{\ssst\rm T} \frac{\delta f}{\delta
  \bar{h}_{ij}}=-2{\rm i}\hbar \bar{K}_{ij}^{\ssst\rm T} \frac{\delta
  D}{\delta \bar{h}_{ij}}\frac{1}{D}f+2 \bar{K}_{ij}^{\ssst\rm T}
\frac{\delta S^{\ssst\rm W}_{1}}{\delta \bar{h}_{ij}}f,\\ 
& & 
-{\rm i}\hbar \left(\,^{\ssst (3)}\! R_{ij}^{\ssst\rm T}+D_{ij}^{\ssst\rm
    T}\right) \frac{\delta f}{\delta \bar{K}_{ij}^{\ssst\rm
    T}}=-{\rm i}\hbar \left(\,^{\ssst (3)}\! R_{ij}^{\ssst\rm
    T}+D_{ij}^{\ssst\rm T}\right)\frac{\delta D}{\delta
  \bar{K}_{ij}^{\ssst\rm T}}\frac{1}{D}f+\left(\,^{\ssst (3)}\!
  R_{ij}^{\ssst\rm T}+D_{ij}^{\ssst\rm T}\right)\frac{\delta
  S^{\ssst\rm W}_{1}}{\delta \bar{K}_{ij}^{\ssst\rm T}}f. 
\end{eqnarray*}
These expressions are used in (\ref{eqn:W-sc0}) to eliminate the second,
third and fourth terms, after multiplying with $f$. As a result, one
obtains 
\begin{eqnarray}
\label{eqn:W-scSchD}
& & \frac{{\rm i}}{2}\bar{h}_{ik}\bar{h}_{jl}\frac{\delta S^{\ssst\rm
    W}_{0}}{\delta \bar{K}_{ij}}\frac{\delta f}{\delta
  \bar{K}_{kl}^{\ssst\rm T}}-{\rm i}\hbar \left(\,^{\ssst (3)}\!
  R_{ij}^{\ssst\rm T}+D_{ij}^{\ssst\rm T}\right)\frac{\delta f}{\delta
  \bar{K}_{ij}^{\ssst\rm T}}-2{\rm i}\hbar \bar{K}_{ij}^{\ssst\rm
  T}\frac{\delta f}{\delta
  \bar{h}_{ij}}+\hat{\mathcal{H}}^{\rm m}_{\bot}f\nonumber\\ 
& & +\left( \frac{{\rm i}}{2}\bar{h}_{ik}\bar{h}_{jl}\frac{\delta^2 S^{\ssst\rm
      W}_{0}}{\delta \bar{K}_{ij}\delta
    \bar{K}_{kl}}-\frac{{\rm i}}{2}\bar{h}_{ik}\bar{h}_{jl}\frac{\delta
    S^{\ssst\rm W}_{0}}{\delta \bar{K}_{ij}^{\ssst\rm T}}\frac{\delta
    D}{\delta \bar{K}_{kl}^{\ssst\rm T}}\frac{1}{D}\right. \nonumber\\
  & & \ \ \left. +{\rm i}\hbar
  \left(\,^{\ssst (3)}\! R_{ij}^{\ssst\rm T}+D_{ij}^{\ssst\rm
      T}\right)\frac{\delta D}{\delta \bar{K}_{kl}^{\ssst\rm
      T}}\frac{1}{D}+2{\rm i} \bar{K}_{ij}^{\ssst\rm T}\frac{\delta
    D}{\delta \bar{h}_{kl}}\frac{1}{D}\right)f =0\,, 
\end{eqnarray}
where $\hat{\mathcal{H}}^{\rm m}_{\bot}f$ comes from
$\left(\left(\hat{\mathcal{H}}^{\rm m}_{\bot}\Psi\right)/\Psi\right)^{(0)}f$. 
We now choose $D$ such that the term in the parenthesis vanishes. This
gives us the equation that defines $D$, in analogy to the situation in
quantum GR \cite{honnef}: 
\[
\label{eqn:W-D}
\frac{{\rm i}}{2}\bar{h}_{ik}\bar{h}_{jl}\frac{\delta^2 S^{\ssst\rm
    W}_{0}}{\delta \bar{K}_{ij}\delta
  \bar{K}_{kl}}-\frac{{\rm i}}{2}\bar{h}_{ik}\bar{h}_{jl}\frac{\delta
  S^{\ssst\rm W}_{0}}{\delta \bar{K}_{ij}^{\ssst\rm T}}\frac{\delta
  D}{\delta \bar{K}_{kl}^{\ssst\rm T}}\frac{1}{D}+{\rm i}\hbar
\left(\,^{\ssst (3)}\! R_{ij}^{\ssst\rm T}+D_{ij}^{\ssst\rm
    T}\right)\frac{\delta D}{\delta \bar{K}_{kl}^{\ssst\rm
    T}}\frac{1}{D}+2{\rm i} \bar{K}_{ij}^{\ssst\rm T}\frac{\delta D}{\delta
  \bar{h}_{kl}}\frac{1}{D}=0 .
\]
With this condition, (\ref{eqn:W-scSchD}) reduces to
\begin{equation}
{\rm i}\hbar\left[-\frac{1}{2\hbar}\bar{h}_{ik}\bar{h}_{jl}\frac{\delta
    S^{\ssst\rm W}_{0}}{\delta \bar{K}_{ij}^{\ssst\rm T}}\frac{\delta }{\delta
    \bar{K}_{kl}^{\ssst\rm T}}+\left(\,^{\ssst (3)}\! R_{ij}^{\ssst\rm
      T}+D_{ij}^{\ssst\rm T}\right)\frac{\delta }{\delta
    \bar{K}_{ij}^{\ssst\rm T}}+2\bar{K}_{ij}^{\ssst\rm T}\frac{\delta
  }{\delta \bar{h}_{ij}}\right]f=\hat{\mathcal{H}}^{\rm m}_{\bot}f. 
\end{equation}
Introducing a local ``bubble'' (Tomonaga-Schwinger) time functional by
\begin{equation}
\label{eqn:WWKBtime}
\frac{\delta}{\delta \tau_{\ssst\rm
    W}(\vec{x})}:=-\frac{1}{2\hbar}\bar{h}_{ik}\bar{h}_{jl}\frac{\delta S_0^{\ssst\rm
    W}}{\delta \bar{K}^{\ssst\rm T}_{kl}}\frac{\delta}{\delta
  \bar{K}^{\ssst\rm T}_{ij}}+\left(\,^{\ssst (3)}\! R_{ij}^{\ssst\rm
    T}+D_{i}D_{j}\right)\frac{\delta}{\delta \bar{K}^{\ssst\rm
    T}_{ij}}+2\bar{K}^{\ssst\rm T}_{ij}\frac{\delta}{\delta
  \bar{h}_{ij}}\,, 
\end{equation}
we arrive at the Tomonaga-Schwinger equation
\begin{equation}
\label{eqn:W-Wtime}
{\rm i}\hbar\frac{\delta f}{\delta
  \tau_{\ssst\rm W}}=\hat{\mathcal{H}}^{\rm m}_{\bot}f\,. 
\end{equation}
Note that $\tau_{\ssst\rm W}$ is, like its GR-counterpart (\ref{tauGR}), not a
scalar function \cite{GK95}. We emphasize that the wave function $f$
is conformally invariant.  

At a formal level, the Tomonaga-Schwinger equation (\ref{eqn:W-Wtime})
resembles the corresponding equation in quantum GR. We see, however,
from the explicit expression for the WKB time (\ref{eqn:WWKBtime})
that it is defined only from the semiclassical shape degrees of
freedom, since the traces (especially $a$) are absent. 
A functional Schr\"odinger equation of the form (\ref{a2.19})
can be derived from the Tomonaga-Schwinger equation by a procedure
similar to the one in GR. This will involve a time parameter that
should be identical with the time parameter of the classical solutions
of Weyl gravity.  

Proceeding with the Born-Oppenheimer scheme to higher orders in 
$\alpha_{\ssst\rm W}$, one arrives at quantum gravitational
corrections terms proportional to $\alpha_{\ssst\rm W}^{-1}$, in
analogy to the higher orders proportional to $m_{\rm P}^{-2}$ in
quantum GR \cite{KS91}. These may serve to study correction terms to
the limit of 
quantum field theory in curved (Weyl) spacetime, but we will not
discuss them here.

\section{Outlook}
\label{sec:4}

Although there is not yet a consensus about the correct quantum theory
of gravity, and about the need to quantize gravity, there exist
several approaches within which concrete questions with potential
observational relevance can be posed and answered. Among them is
canonical quantum gravity in the metric formulation. If general
relativity is quantized in this way, one arrives at the Wheeler-DeWitt
equation and the momentum constraints. A semiclassical expansion leads
to the recovery of quantum field theory in curved spacetime plus
quantum gravitational corrections. The latter may be observationally
tested, for example, in the CMB anisotropy spectrum.

Our concern here was to discuss canonical quantization and the semiclassical
limit for an alternative theory based on the Weyl tensor. This ``Weyl
gravity'' does not contain any scale, so it may be of empirical
relevance only in the early Universe, where scales may be
unimportant. Independent of this possibility, it is of structural
interest to compare this theory in its quantum version with quantum
general relativity. We have seen here that a semiclassical limit can be
performed by a well defined approximation scheme, although the
emerging semiclassical time has properties different from standard
semiclassical time. 
In future investigations, we plan to apply a theory based on the sum
of Weyl and Einstein-Hilbert action
to the early Universe and to the understanding of spacetime structure
at a fundamental level, topics that are also at the centre of Paddy's
interest.


\begin{thebibliography}{99.}%

\bibitem{Bach21} R. Bach, Math. Z. {\bf 9}, 110 (1921)

\bibitem{Barbour12} J. Barbour, in \textit{Quantum Field Theory and
    Gravity}, ed. by F.~Finster {\em et al.} (Springer, Basel, 2012),
  p.~257 

\bibitem{Boulware83} D. Boulware, in \textit{Quantum Theory of
    Gravity}, ed. by S.M. Christensen (Adam Hilger Ltd, Bristol, 1984),
  p. 267

\bibitem{BKK16} D. Brizuela, C. Kiefer, M. Kr\"amer,  Phys. Rev. D
  {\bf 93}, 104035 (2016)

\bibitem{BK96} T. Brotz, C. Kiefer, Nucl. Phys. B {\bf 475}, 339 (1996)
%
\bibitem{DerrHD} N.~Deruelle, M.~Sasaki, Y.~Sendouda,
D.~Yamauchi, Prog. Theor. Phys. {\bf 123}, 169 (2010)

\bibitem{Einstein21} A. Einstein, Preu\ss.
  Akad. Wiss. (Berlin). Sitzungsberichte (1921): 261--264

\bibitem{EinsteinColl} A. Einstein, {\em Collected Papers}, Vol.~{\bf 12} 
(Princeton University Press, Princeton, 2009)

\bibitem{GK95} D.~Giulini, C.~Kiefer, Class. Quantum Grav. {\bf
    12}, 403 (1995) 

\bibitem{Greensite} J. Greensite, Nucl. Phys. B {\bf 351}, 749 (1991)

\bibitem{Goenner} H.F.M. Goenner, Living Rev. Relativity {\bf 7},
\url{http://www.livingreviews.org/lrr-2004-2 . Cited 13 July 2016}

\bibitem{ILP} M.~Irakleidou, I.~Lovrekovic, F.~Preis, Phys. Rev. D
  {\bf 91}, 104037 (2015)

\bibitem{CK87} C. Kiefer, Class. Quantum Grav. {\bf 4}, 1369 (1987)

\bibitem{honnef} C. Kiefer, in \textit{Canonical
    Gravity: From Classical to Quantum}, ed. J.~Ehlers and
  H.~Friedrich (Springer, Berlin, 1994), p. 170

\bibitem{OUP} C. Kiefer, \textit{Quantum Gravity}, 3rd edn. (Oxford
  University Press, Oxford, 2012). 

\bibitem{KN16a} C. Kiefer, B. Nikoli\'c, Conformal and Weyl-Einstein
  gravity I. Classical Geometrodynamics, to be submitted (2016)

\bibitem{KN16b} C. Kiefer, B. Nikoli\'c, Conformal and Weyl-Einstein
  gravity II. Quantum Geometrodynamics, to be submitted (2016)
%
\bibitem{KS91} C. Kiefer, T.P. Singh, Phys. Rev. D \textbf{44}, 1067
  (1991) 
  
  \bibitem{KPS91} C. Kiefer, T. Padmanabhan, T. P. Singh,
    Class. Quantum Grav. {\bf 8}, L185 (1991) 

\bibitem{KOT14} J. Kluso\u{n}, M. Oksanen, A. Tureanu,  Phys. Rev. D
  {\bf 89}, 064043 (2014)
%
\bibitem{Maldacena11} J. Maldacena, Einstein Gravity from Conformal
  Gravity, arXiv:1105.5632v2 [hep-th] 

\bibitem{Mannheim12} P.D. Mannheim, Found. Phys. {\bf 42}, 388 (2012)

\bibitem{Paddy85} T. Padmanabhan, Ann. Phys. (N.Y.) {\bf 165}, 38 (1985)


\bibitem{Paddy89} T. Padmanabhan, Phys. Rev. D {\bf 39}, 2924 (1989)

\bibitem{Paddy90}  T. Padmanabhan, Pramana - J. Phys. {\bf 35}, L199 (1990)

\bibitem{PX95} V. Perlick and C. Xu, Astrophys. J. {\bf 449}, 47
  (1995)

 \bibitem{Pitts14} B. Pitts, Ann. Phys. (N.Y.) {\bf 351}, 382 (2014) 
  
\bibitem{Schmidt06} H.-J. Schmidt, Int. J. Geom. Methods Mod. Phys
  {\bf 04}, 209 (2007)

  \bibitem{SP89} T. P. Singh, T. Padmanabhan, Ann. Phys. (N.Y.) {\bf 196}, 296 (1989)
  
\bibitem{Thiemann} T. Thiemann, {\em Modern canonical 
           quantum general relativity} (Cambridge University Press,
           Cambridge, 2007)

\bibitem{Hooft15} G. 't Hooft, Int. J. Mod. Phys. D {\bf 24}, 1543001 (2015)

 \bibitem{Zeh86} H. D. Zeh, Phys. Lett. A {\bf 116}, 9 (1986)

\end{thebibliography}
\end{document}